\documentclass[%
 reprint,
 amsmath,amssymb,
 aps,
]{revtex4-1}

\usepackage{graphicx}% Include figure files
\usepackage{dcolumn}% Align table columns on decimal point
\usepackage{bm}% bold math
\usepackage{braket}
\usepackage{float}

\newcommand{\Tc}{$T_\mathrm{c}$ }
\newcommand{\Jc}{$J_\mathrm{c}$ }

\newcommand{\Jcm}{J_\mathrm{c} }
\graphicspath{{./fig-intro/}}
\newcommand{\YBCO}{YBa$_2$Cu$_3$O$_{7-\delta}$ }
\newcommand{\DBCO}{DyBa$_2$Cu$_3$O$_{7-\delta}$ }
\newcommand{\BaKx}{Ba$_{1-x}$K$_x$Fe$_2$As$_2$ }
\newcommand{\BaKxn}{Ba$_{1-x}$K$_x$Fe$_2$As$_2$}
\newcommand{\BaPx}{BaFe$_2$(As$_{1-x}$P$_x$)$_2$ }
\newcommand{\BaCox}{Ba(Fe$_{1-x}$Co$_x$)$_2$As$_2$ }

\newcommand{\Tcm}{T_\mathrm{c}}
\newcommand{\Deg}{^\circ}

\begin{document}

\preprint{APS/123-QED}

\title{Field-driven transition in the Ba$_{1-x}$K$_x$Fe$_2$As$_2$ superconductor \\
 with splayed columnar defects}% Force line breaks with \\

\author{Akiyoshi Park}
\author{Sunseng Pyon}
\author{Kengo Ohara}
\author{Nozomu Ito}
\author{Tsuyoshi Tamegai}
\affiliation{Department of Applied Physics, The University of Tokyo, Hongo, Bunkyo-ku, Tokyo 113-8656, Japan\\}

\author{Tadashi Kambara}
\author{Atsushi Yoshida}
\affiliation{Nishina Center, RIKEN, 2-1 Hirosawa, Wako, Saitama 351-0198, Japan}

\author{Ataru Ichinose}
\affiliation{Central Research Institute of Electric Power Industry, Electric Power Engineering Research Laboratory, 2-6-1, Nagasaka, Yokosuka-shi, Kanagawa, Japan}

\date{\today}

\begin{abstract}
Through 2.6 GeV U irradiations, we have induced bimodal splayed columnar defects in \BaKx single crystals with splay angles, $\pm 5 \Deg$, $\pm 10 \Deg$, $\pm 15 \Deg$, and  $\pm 20 \Deg$.  Critical current densities through magnetization measurements were carefully evaluated, where a splay angle of $\pm 5 \Deg$ brought about the highest $\Jcm$.  Mageto-optical images close to \Tc indicates highly anisotropic discontinuity lines in the remnant state, and with anisotropy increasing with greater splay angles. 

Moreover, amongst those with splayed columnar defects, anomalous non-monotonic field dependences of \Jc and $S$ with an extrema at some fraction of the matching field are observed. We discuss that such \Jc enhancement arises from a field-driven coupling transition in which intervortex interactions reorganize the vortex structure to be accommodated into columnar defects, thereby increasing pinning at higher fields.

\begin{description}

\item[PACS numbers]
%74.70.Xa, 74.62.En, 74.25.fc
74.70.Xa, 74.25.Sv, 74.25.Wx
\end{description}
\end{abstract}
\maketitle

%%%%%%%%%%%%%%%%%%%%%%%%%%%%%%%%%%%%%%%%%%%%%%%%
\section{Introduction}
Motion of flux lines in the mixed state of type-II superconductors has a detrimental consequence of impairing its dissipation-less `zero dc resistivity' state.  Retaining stability of flux lines has therefore been a challenge as it is a matter of its high technological interest.  As a remedy to such a problem, the notion of localizing flux lines within parallel tracks of columns was originally suggested by portraying the highly localized vortex phase as a Bose glass \cite{PhysRevB.48.13060}.  Such a remarkable enhancement of pinning was confirmed experimentally through observing remarkable increase in the critical current density ($\Jcm$) in cuprate \cite{PhysRevLett.67.648} and iron-based superconductors (IBSs) \cite{PhysRevB.80.012510, PhysRevB.81.094509, doi:10.1063/1.4731204, Fang2013, doi:10.1063/1.4829524, 0953-2048-28-5-055011, 0953-2048-25-8-084008} after incorporating columnar defects via heavy-ion irradiation.

\begin{figure}[t]
  \begin{center}
\includegraphics[width=8.5 cm]{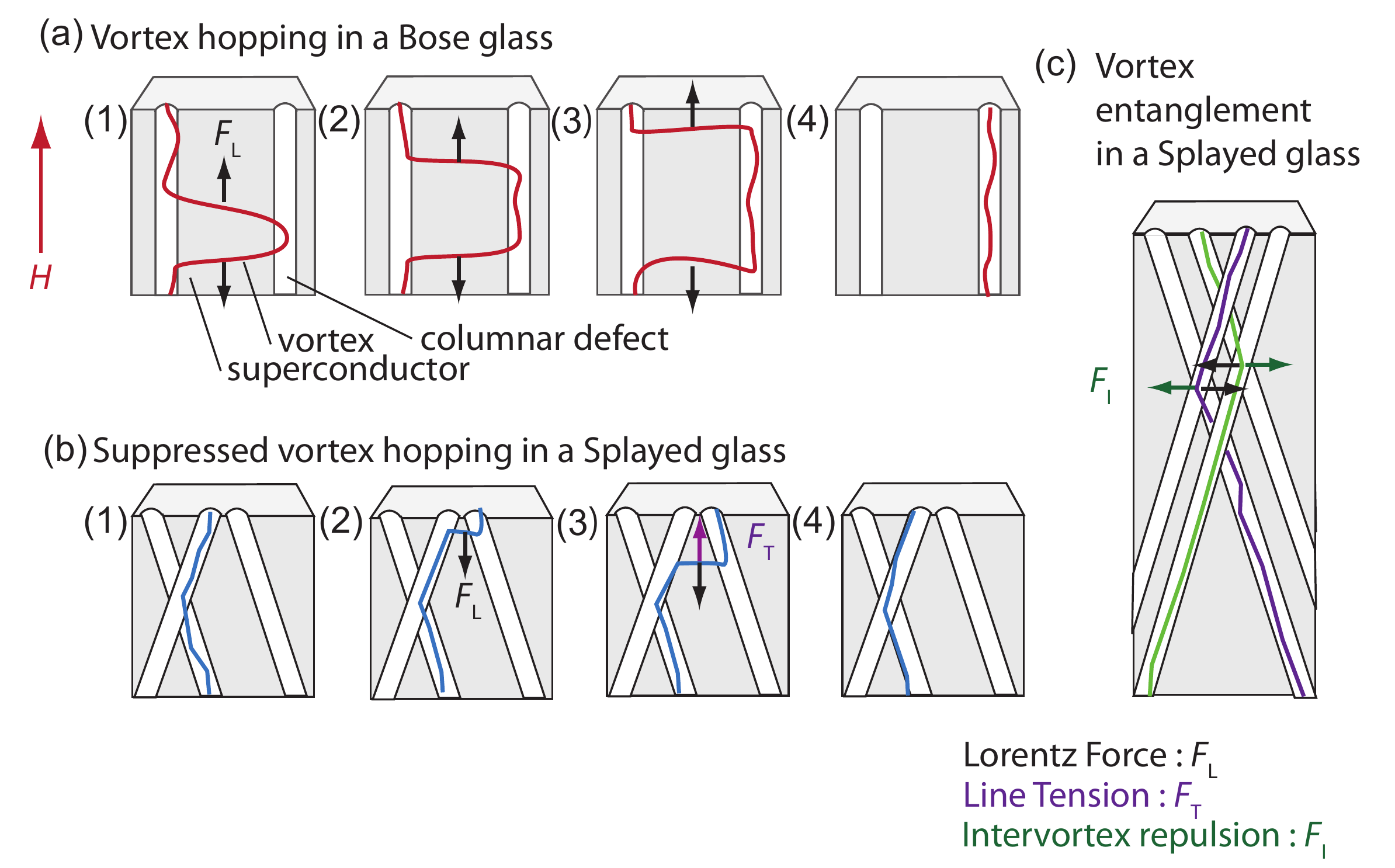}% Here is how to import EPS art
  \end{center}
\caption{ \footnotesize{(a) Model of vortex hopping from one column to another in a Bose glass phase. (b) Reduction of vortex hopping caused by variable inter-defect range in a splayed glass phase.  (c)  Flux entanglement due to intersecting columnar defects in a splayed glass phase.}}
\label{vortex}
\end{figure}

Later, further enhancement of $\Jcm$ by dispersing the angles of columnar defects was suggested by Hwa \textit{et al.} \cite{PhysRevLett.71.3545}. As illustrated in Fig. \ref{vortex}(a), for the case of Bose glass phase in which columnar tracks are parallel, thermal activation may prompt a segment of the flux line to extend to a neighboring defect, allowing the rest of the flux to relocate itself without any expenditure of energy, ultimately leading to hopping. On the other hand, for splayed columnar defects, the variable inter-defect distance makes relocation of vortex through thermal activation energetically unfavorable, thereby strongly suppressing vortex motion as shown in Fig. \ref{vortex}(b). Moreover, the splayed columnar defects may promote forced entanglement of vortices, additionally enhancing $\Jcm$ \cite{PhysRevLett.71.3545}.  Nonetheless, tilting the columnar defects above the lock-in angle is inimical to flux pinning, as pinning is most robust when aligned to the applied field \cite{PhysRevB.50.4102}.  Such an inherent competition between the adverse effect of vortex-field misalignment and beneficial effect of splaying columnar defects raises a question: which splay angle optimally enhances the $\Jcm$.  Current knowledge concerning the optimal splay angle is limited to a seminal report on Au-irradiated \YBCO crystals in which a splay angle of $\pm 5 \Deg$ yielded the largest $\Jcm$ amongst   $\pm 0 \Deg, \pm 5 \Deg, \pm 10 \Deg,$ and $\pm 15 \Deg$  \cite{PhysRevLett.76.2563}.  Similar results were also indicated by Park \textit{et al.} in 1.3 GeV U irradiated \YBCO thin films \cite{PARK1997310}, and as well as in 6 GeV irradiated \YBCO crystals  \cite{PhysRevB.59.8455}.
% Not only in cuprate superconductors, even in 2.6 GeV U irradiated Ba$_{0.6}$K$_{0.4}$Fe$_2$As$_2$, a splay angle of $\pm 5 \Deg$ was reported to reveal the highest self field \Jc at 2 K \cite{PARK201658}.
Yet, the effects of larger splay angles and the effects of splaying columnar defects amongst other superconducting systems with differing vortex structures yet remain nebulous.  Understanding the role of splayed columnar defects in IBSs cultivates an insight into designing improved pinning landscape for serving the best of our purpose. 

Previously, in Ref. \cite{PARK201658} we have explored the effects of incorporating bimodal splayed columnar defects in IBSs through irradiating optimally doped \BaKx with 2.6 GeV $^{238}$U ions and provided evidence that a splay angle of $\pm 5 \Deg$ yields the largest self-field $\Jcm$ at 2 K.  Through this article, we confirm the same trend through a different set of samples, and  
%change in the critical current densities in by varying the splay angle through irradiating optimally doped \BaKx with 2.6 GeV $^{238}$U ions \cite{PARK201658}.
%Here, we report the effects of incorporating bimodal splayed columnar defects in IBSs by irradiating optimally doped \BaKx with 2.6 GeV $^{238}$U ions and exploring the resulting change in vortex dynamics and critical current densities.
%One key results is that we 
reveal via magneto-optical imaging, the presence of two components in the in-plane $\Jcm$: a component perpendicular to the splay plane ($\Jcm^{\perp\mathrm{ splay}}$) and a parallel component ($\Jcm^{||\mathrm{ splay}}$), in which $\Jcm^{||\mathrm{ splay}}>\Jcm^{\perp\mathrm{ splay}}$ at high temperatures.  Despite the large anisotropy revealed close to $\Tcm$, we discuss that the anisotropy reduces to unity at low temperatures, thus allowing us to compare the self-field $\Jcm$ value at 2 K between different samples without having to quantify the individual $\Jcm$ components.
%Furthermore, there is an apparent increase in the anisotropy with larger splay angles, especially at higher fields and temperatures.  
%Although we have calculated the overall $\Jcm$ though magnetization measurements, 
%We discuss that such presence of anisotropy is an indication that the $\Jcm^{||\mathrm{ splay}}$ is actually greater than the overall $\Jcm$ quantified through magnetization measurements, thereby further affirming the robustness of pinning by splayed columnar defects.%, compared to systems with those with conventional parallel columnar defects.
% 
%Secondly, 
Another key results is that we observe a novel vortex phenomena that brings about an anomalous secondary magnetization peak in the magnetic hysteresis curve when the magnetic field is applied along the average direction of the splayed columnar defects.  We provide evidence that such non-monotonicity is a result of vortex-vortex interactions accommodating flux into columnar defects at higher fields, thereby enhancing pinning.

%%%%%%%%%%%%%%%%%%%%%%%%%%%%%%%%%%%%%%%%%%%%%%%%
\section{Experimental Details}
\begin{figure}[t]
  \begin{center}
\includegraphics[width=8.7 cm]{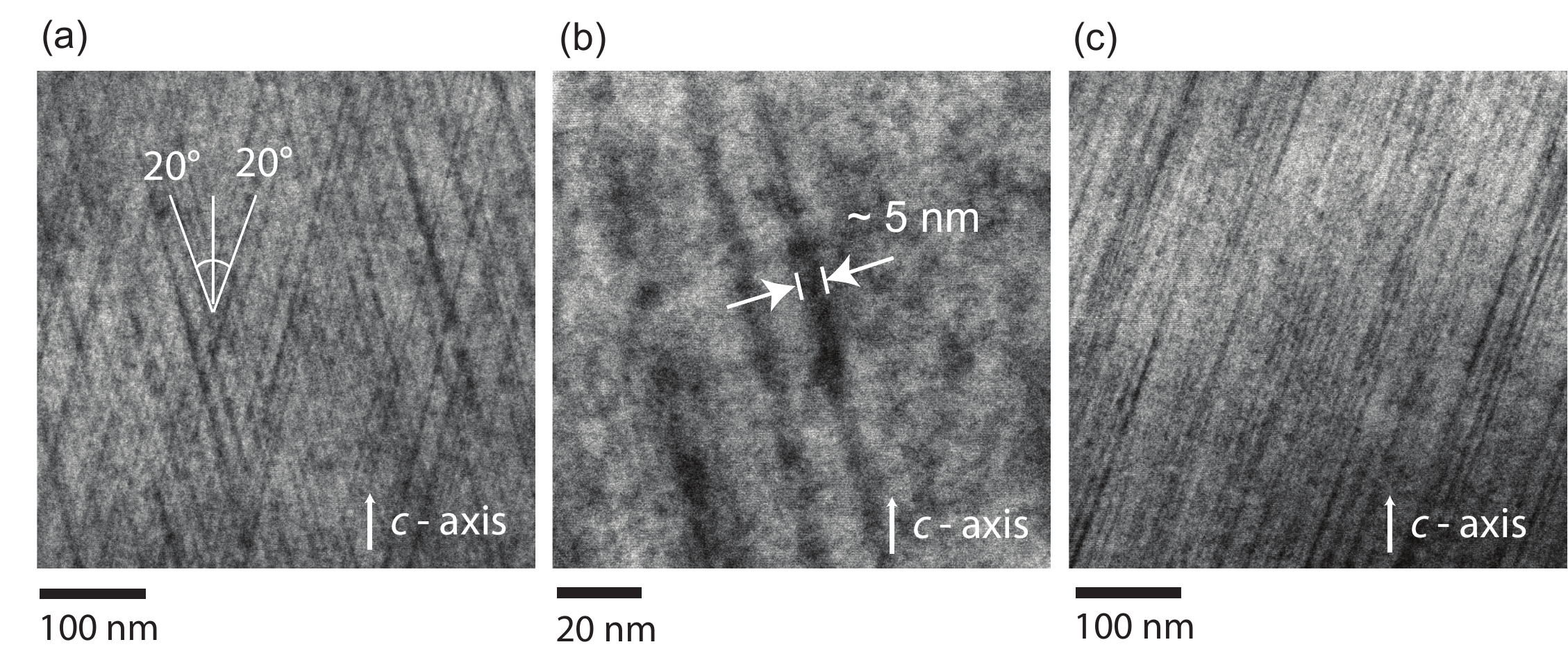}% Here is how to import EPS art
  \end{center}
\caption{ {Transmission electron micrographs of cross-sections of 2.6 GeV U irradiated Ba$_{1-x}$K$_x$Fe$_2$As$_2$.  (a)  \BaKx irradiated with a total dose of 8 T (4T $+$ 4 T) with a splay angle of $\pm 20 \Deg$.  Note that the angle shown in the micrographs may appear to be smaller due to slight deviation of the observed cross-sectional plane from the splay plane.  (b) A zoomed in cross-sectional micograph of the sample in (a) indicating the diameter of the columnar defect. (c) \BaKx with a total dose of 8 T with tilted columnar defect of angle $20 \Deg$ from the $c$-axis.}}
\label{splaytem}
\end{figure}

For this experiment, \BaKxn, a prototypical IBS was employed for investigation.   With optimal doping, the \Tc reaches 38 K, the highest amongst the BaFe$_2$As$_2$ system.  Moreover, the small coherence length $\xi_0 = 1.2$ nm \cite{PhysRevB.79.094505} in optimally doped \BaKx compared to optimally doped \BaPx with  $\xi_0 = 2.14$ nm \cite{PhysRevB.85.184525} and \BaCox with  $\xi_0 = 2.44$ nm \cite{:/content/aip/journal/apl/94/6/10.1063/1.3081455} indicates that \BaKx has a substantially higher condensation energy ($\varepsilon_0/4\xi^2$, where $\varepsilon_0$ is the line energy) amongst others, thereby making core pinning prompted by artificial defects to be much more effective.  The highest reported enhancement of critical current density in \BaKx has been achieved through 320 MeV Au and 2.6 GeV U irradiation \cite{Ohtake201547}.  Hence, in pursuing a high critical current density through sculpting the most effective pinning landscape, \BaKx would be an excellent target material. 

Here, \BaKx single crystals were grown through a FeAs flux method.  Nominal amounts of Ba : K : FeAs were put with a ratio of $1\hspace{-1 pt}-\hspace{-1 pt}x : 1.1x : 4$ into an alumina crucible.  For the present case, the optimal doping level $x = 0.40$ was employed.  Specifically, Ba plates and K chunks together with FeAs powder were placed inside an alumina crucible in a N$_2$ atmosphere glove box, then sealed inside a stainless steel tube with a stainless steel cap \cite{doi:10.1143/JPSJ.79.124713}.  The reason why a stainless steel seal was employed is because quartz is understood to react with K, making the quartz brittle, upon heating.  The assembly was heated up to 1150 $\Deg$C over a period of 10 hours and cooled to 800 $\Deg$C over a period of 70 hours, then finally furnace cooled to room temperature \cite{doi:10.1143/JPSJ.79.124713}.  Within the flux, crystal platelets with dimensions over $1 \times 1 \times 0.05$ mm$^3$ were retrieved.  Energy Dispersive X-ray (EDX) spectroscopy analysis affirmed homogeneous doping of $x=0.40$, and magnetization and resistivity measurements revealed a \Tc of 38.6 K.  The crystals were cleaved into a rectangular geometry and subject to irradiation.

Uranium irradiation was performed at the RIKEN Nishina Center with $^{238}$U ions with energy
of 10.75 MeV per nucleon, which translates to 2.6 GeV per ion.  The ions were irradiated at room temperature, assuming that annealing of defects do not take place.  Moreover, all samples were irradiated with a total matching field of $B_\Phi=8$ T.  Once irradiation was performed, samples were subject to magnetization measurements and magneto-optical (MO) imaging.

%%%%%%%%%%%%%%%%%%%%%%%%%%%%%%%%%%%%%%%%%%%%%%%%
\section{Results}

\begin{figure}[h!]
  \begin{center}
\includegraphics[width=8 cm]{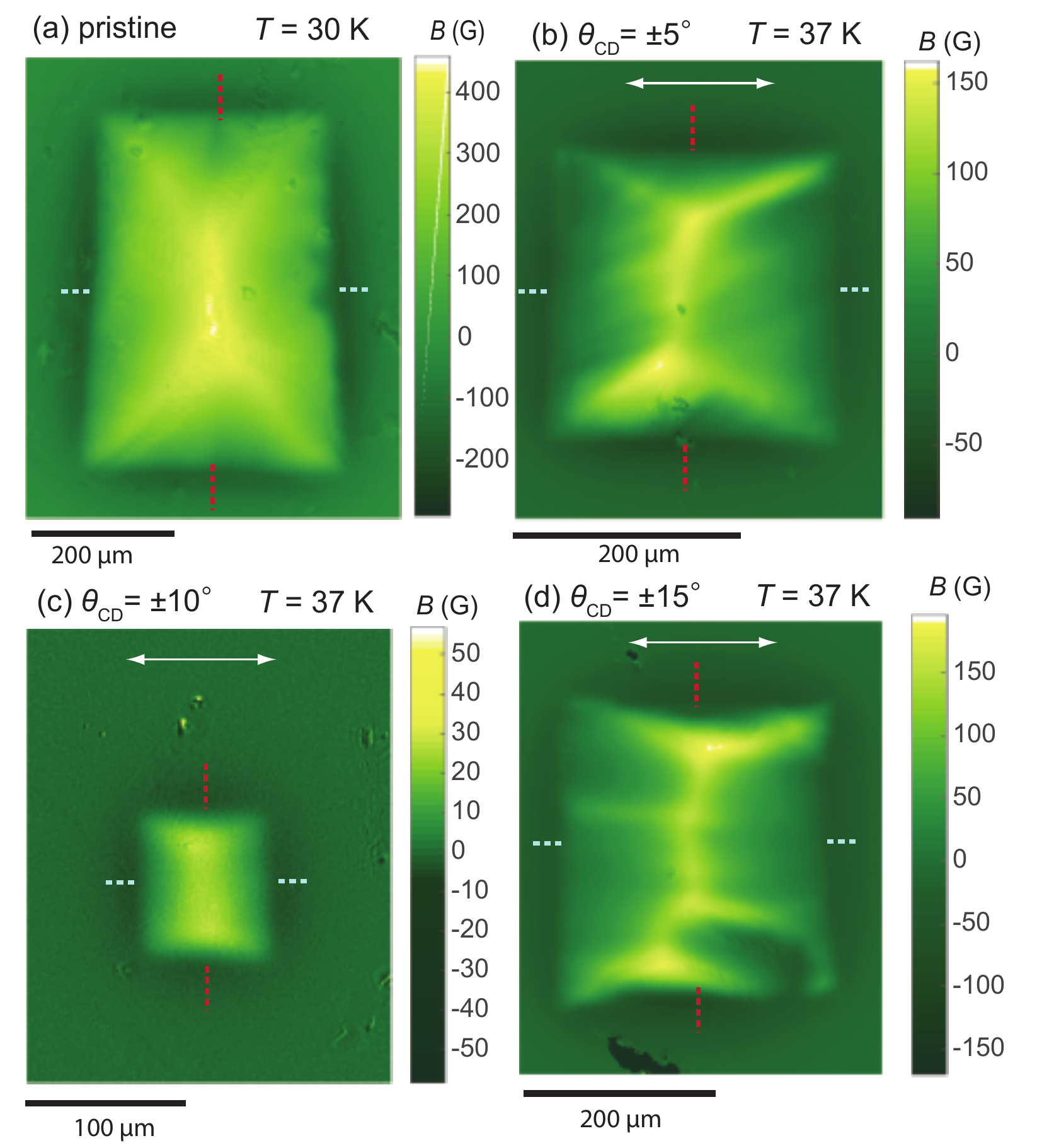}% Here is how to import EPS art
  \end{center}
\caption{ {Magneto-optical images of \BaKx single crystal (a) in the pristine state at $T = 30$ K, and those irradiated with splay angle of (b) $\pm 5 \Deg$, (c) $\pm 10 \Deg$ and (d) $\pm 15 \Deg$ in the remnant state using a field of 1 kOe along the $c$-axis, at $T=37$ K.  The white arrows show the splay direction.  Furthermore, the red and white dashed lines depict where the line-profiles are extracted.} }
\label{M05}
\end{figure}
\begin{figure*}[t]
  \begin{center}
\includegraphics[width=\textwidth]{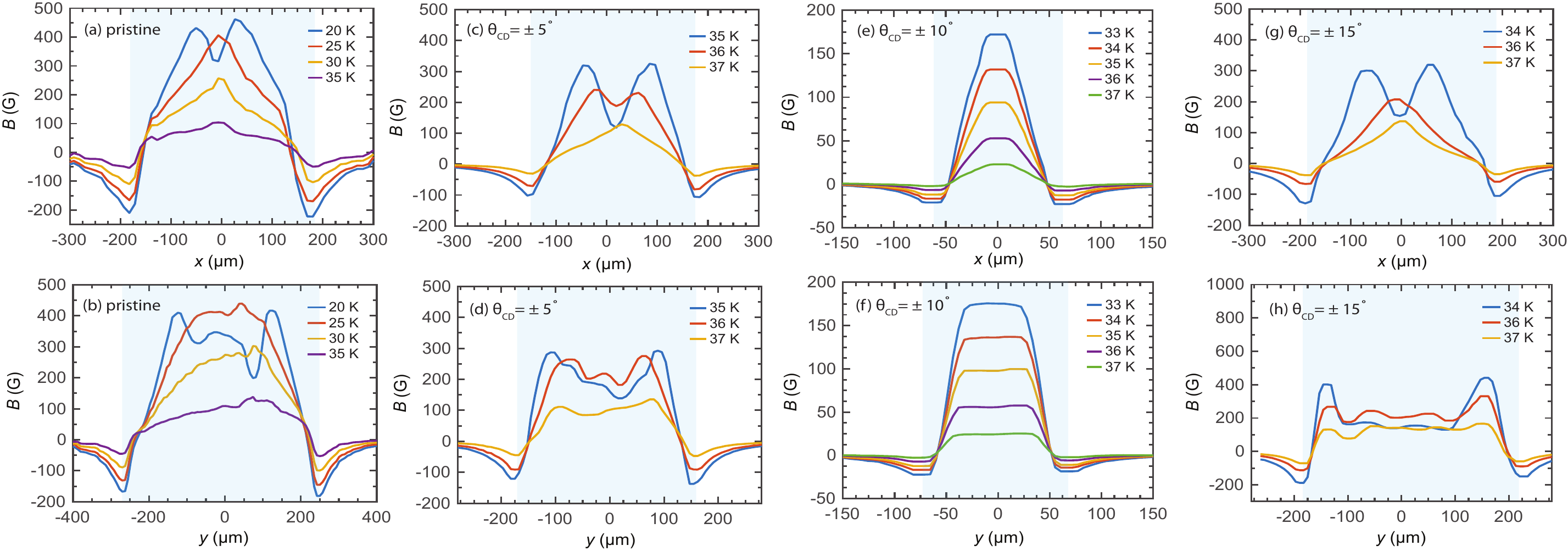}% Here is how to import EPS art
  \end{center}
%\vspace{-100 pt}
\caption{ {Line profiles of MO images of  Fig. \ref{M05} along the white dashed lines (a), (c), (e), (g) and along the red dashed lines  (b), (d), (f), (h) at various temperatures.  The blue regions indicate the width of the sample.} }
\label{M15}
\end{figure*}

\begin{figure}[b]
  \begin{center}
\includegraphics[width=7.5 cm]{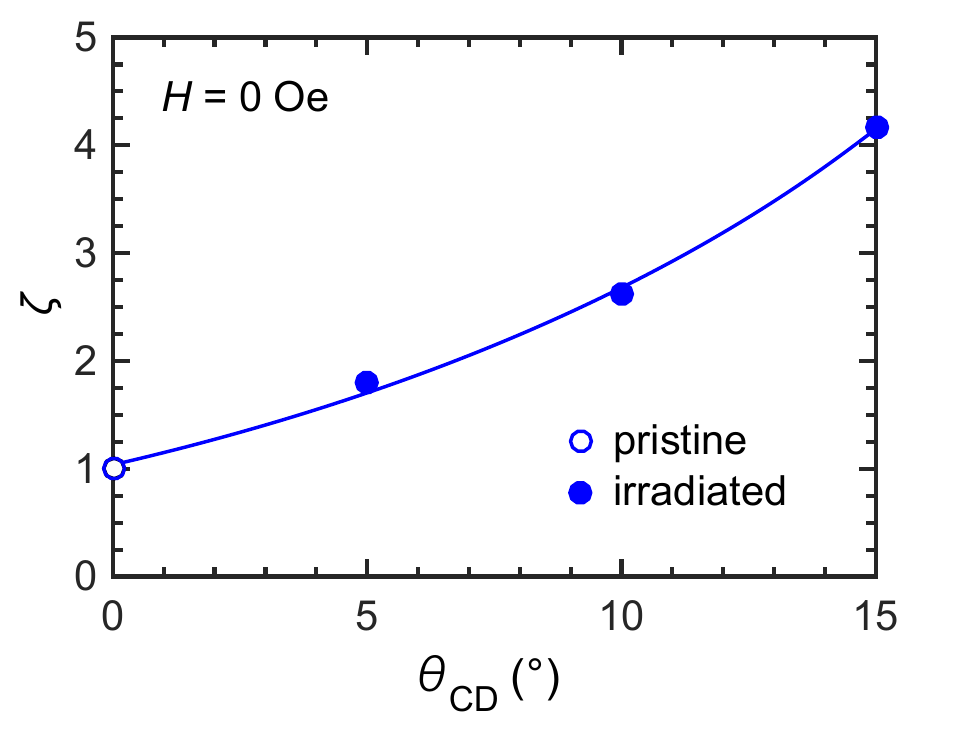}% Here is how to import EPS art
  \end{center}
\caption{ {The splay angle dependence of $\Jcm$ anisotropy, $\zeta$, calculated from the MO images of Fig. (\ref{M05}) close to $\Tcm$ in the remnant state.} }
\label{aniso}
\end{figure}

\subsection{Defect Structure}
For detailed discussions, it is crucial to be aware of the type of defects incorporated in the system. As exhibited in Fig.  \ref{splaytem}(a), 2.6 GeV U irradiation in \BaKx indeed introduces linear tracks of columnar defects that cross each other.  Opposed to 320 MeV Au irradiation which produces segmented columnar defects \cite{0953-2048-25-8-084008}, 2.6 GeV U irradiation introduces continuous columnar defects, which makes the splay more effective.  The diameter of each column is about 3 - 6 nm (Fig.  \ref{splaytem}(b)), comparable to the scale of the coherence length. Evidently, the similar case is seen for those irradiated with tilted columnar defects (Fig.  \ref{splaytem}(c)). The size makes each of the columns excellent pinning center for core interaction.  Based on the morphology of the defects elucidated through scanning transmission electron microscope (STEM) observation, the physics of vortex matter will be discussed here on.

\begin{figure*}
  \begin{center}
\includegraphics[width=\textwidth]{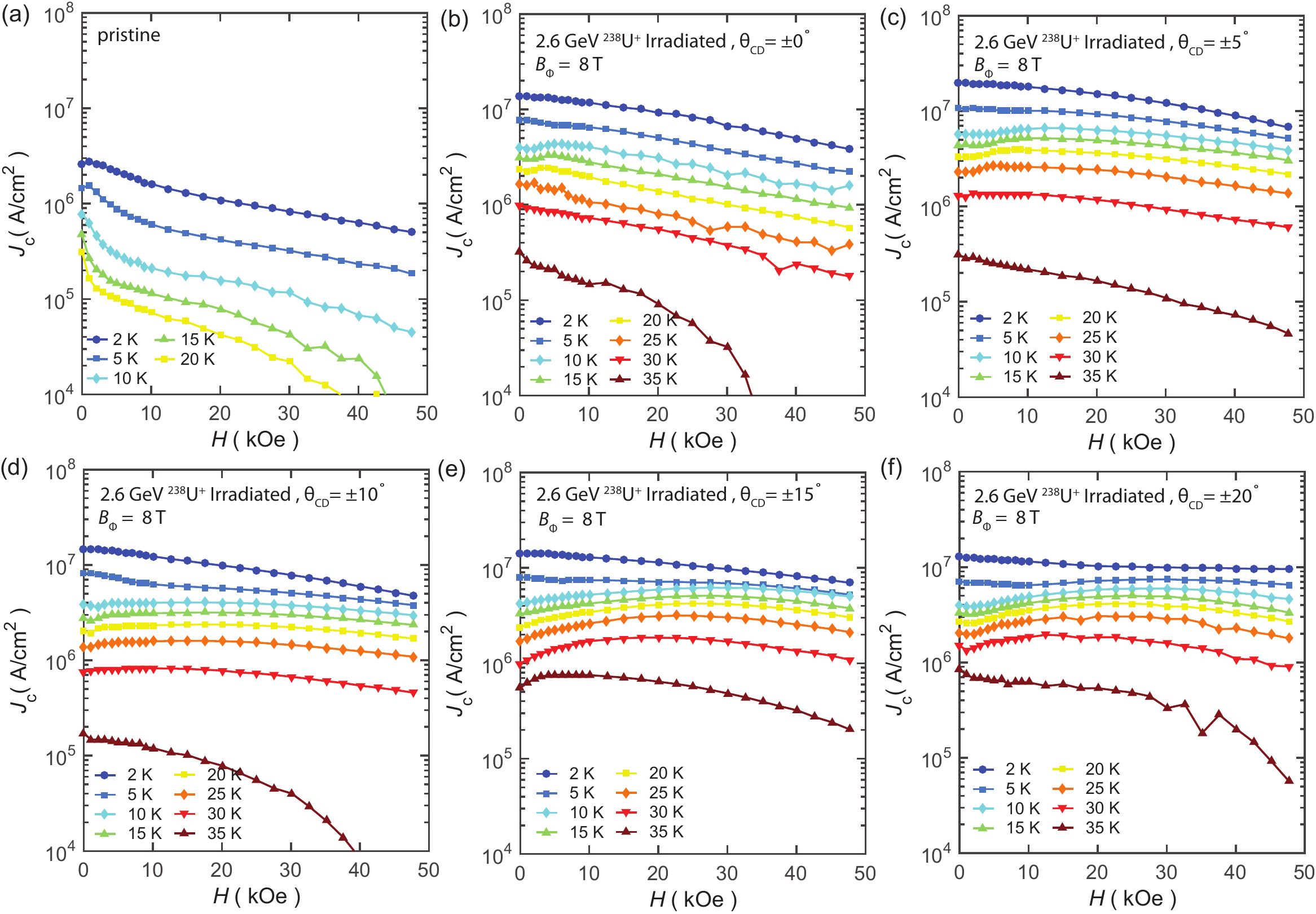}% Here is how to import EPS art
  \end{center}
\caption{ {Magnetic field dependence of $\Jcm$ in \BaKx of (a) pristine sample and those after $^{238}$U irradiation with (b) parallel defects and splay angle of (c) $\pm 5 \Deg$, (d) $\pm 10 \Deg$, (e) $\pm 15 \Deg$, and  (f) $\pm 20 \Deg$.}}
\label{JcH}
\end{figure*}

\begin{figure}[b]
  \begin{center}
\includegraphics[width=8.5 cm]{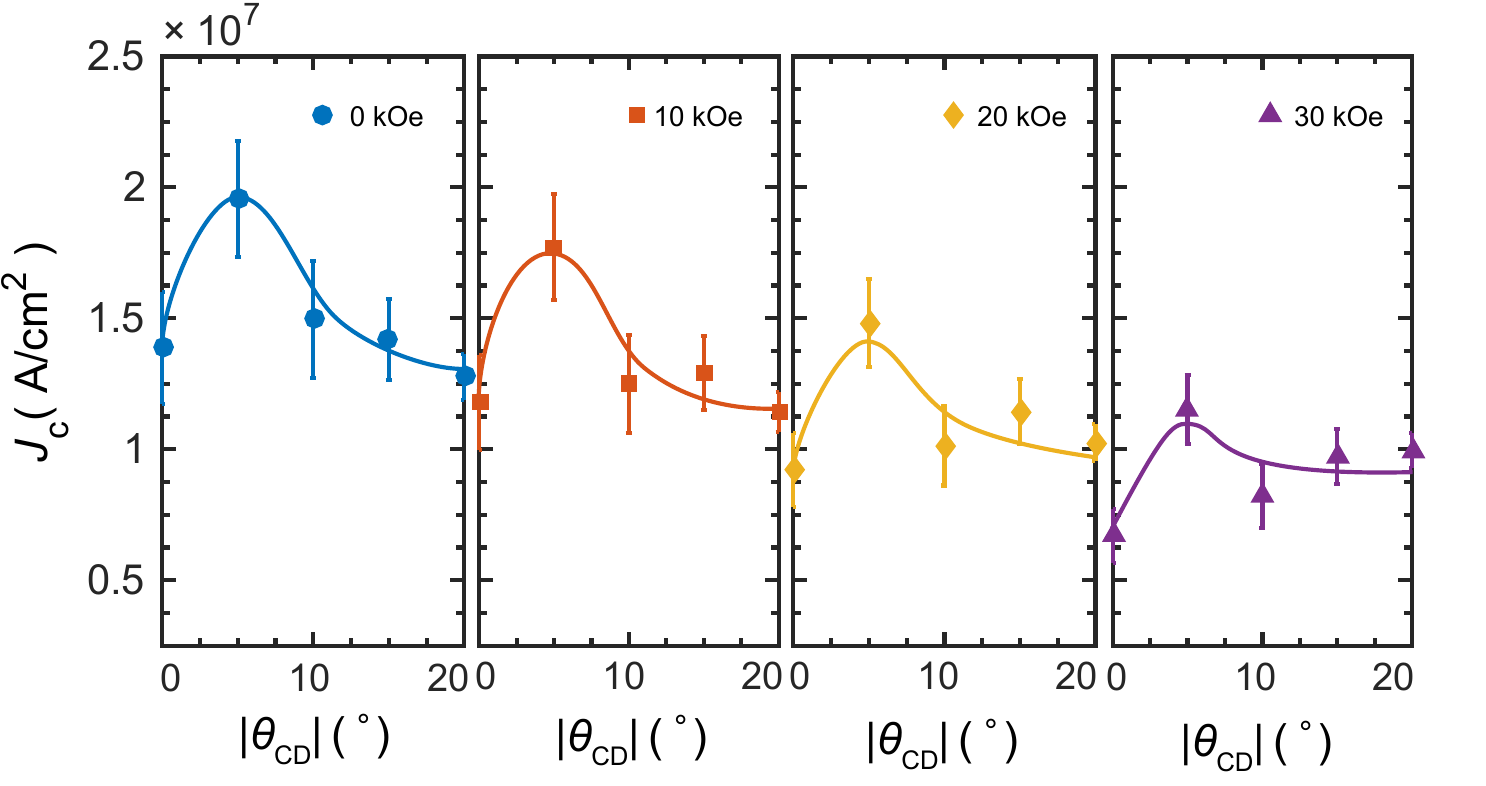}% Here is how to import EPS art
  \end{center}
\caption{ {Splay angle dependence of $\Jcm$ in \BaKx at 2 K under various applied fields. Evidently, the highest \Jc is achieved at small splay angles.  The error bars indicate the possible error in \Jc stemming from the error in the measurement of crystal thickness $c$ with uncertainty of 1.5 $\mu$m.}}
\label{angle}
\end{figure}

\begin{figure*}[t]
  \begin{center}
\includegraphics[width=\textwidth]{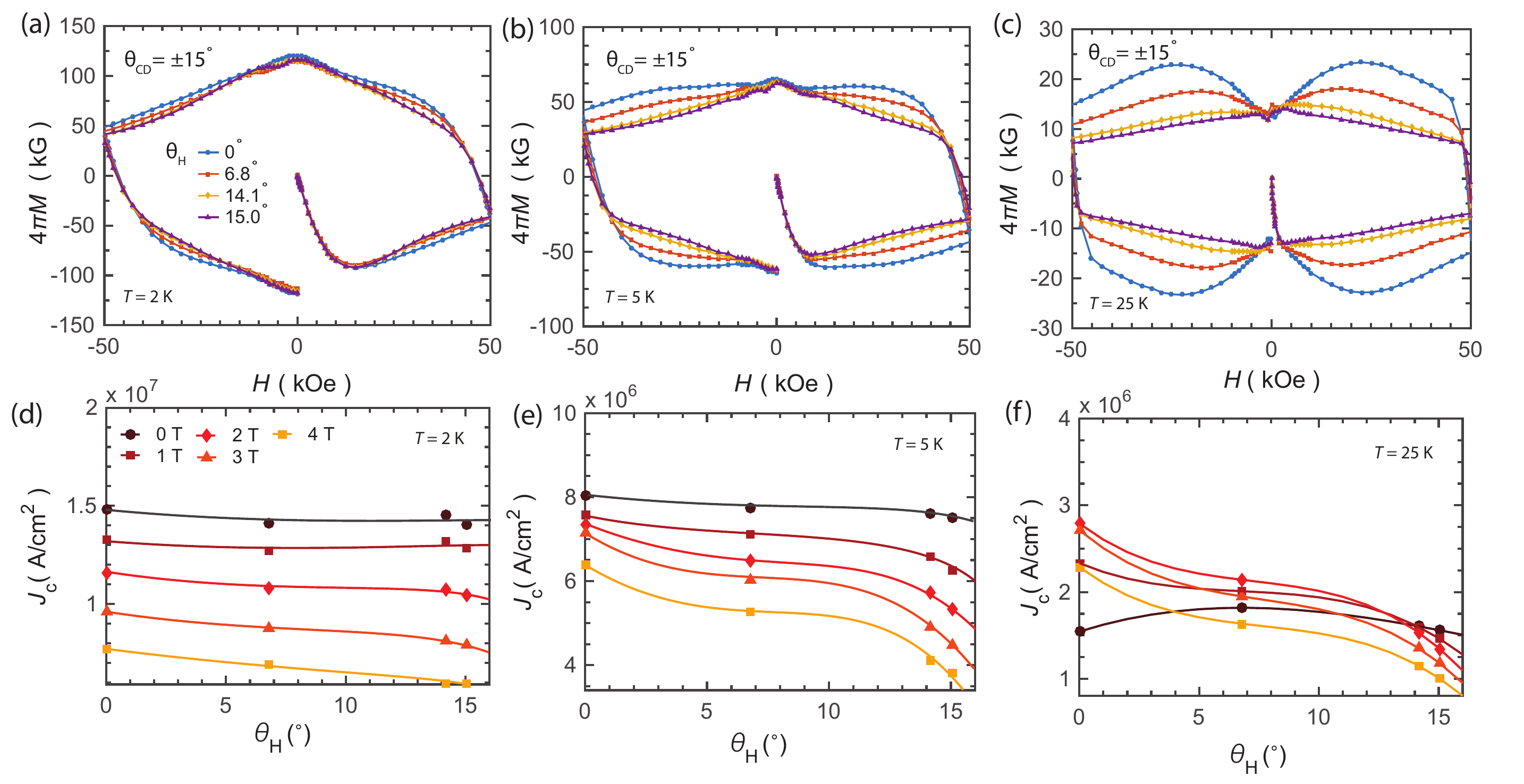}% Here is how to import EPS art
  \end{center}
\caption{ \footnotesize{Magnetic hysteresis loops of \BaKx with bimodal splay columnar defects of $\theta_\mathrm{CD} = \pm 15 \Deg$ with a total dose of $B_\Phi = 8$ T in tilted fields of various angles at (a) 2 K (b) 5 K, and (c) 25 K.  The \Jc dependence of the angle of tilted field at (d) 2 K (e) 5 K, and (f) 25 K.} }
\label{tilt1}
\end{figure*}
\begin{figure*}[t]
  \begin{center}
\includegraphics[width=\textwidth]{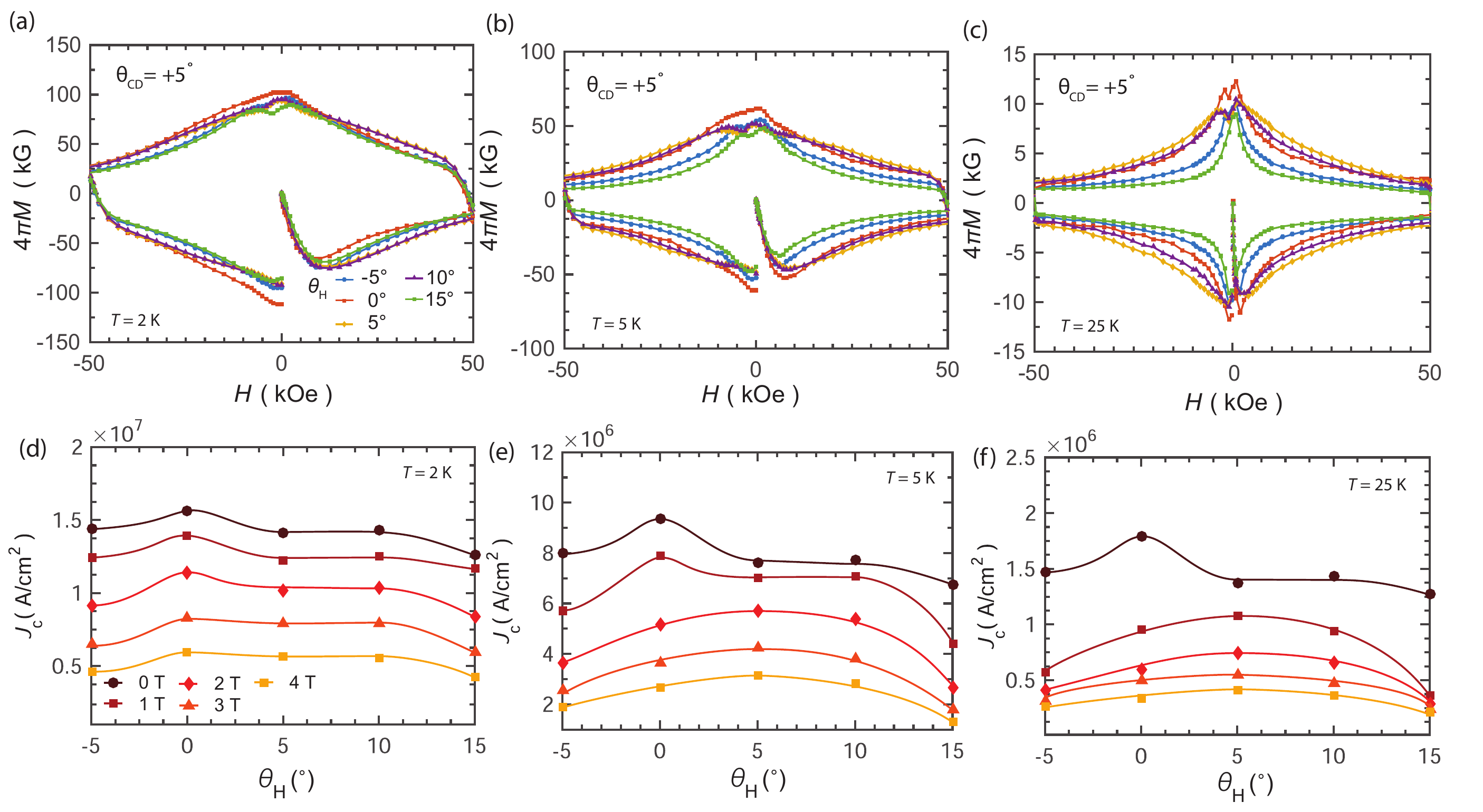}% Here is how to import EPS art
  \end{center}
\caption{ \footnotesize{Magnetic hysteresis loops of \BaKx with tilted columnar defects of $\theta_\mathrm{CD} = 5 \Deg$ with a total dose of $B_\Phi = 8$ T in tilted fields of various angles at (a) 2 K (b) 5 K, and (c) 25 K. The tilt-angle dependence of \Jc at (d) 2 K (e) 5 K, and (f) 25 K.} }
\label{tilt2}
\end{figure*}

\subsection{Anisotropic critical current density}
For a bimodal splay system, as in the case here, two components of \Jc arise : $\Jcm^{||\mathrm{ splay}}$  and  $\Jcm^{\perp\mathrm{ splay}}$, the critical current density that runs in the same direction as the splay plane and critical current density that runs perpendicular to the splay plane.  The existence of two different \Jc component amongst systems with splayed columnar defects has been confirmed via resistivity measurements in 1 GeV Au irradiated \YBCO by Lopez \textit{et al.} and in 3.9 GeV Au irradiated \YBCO by Kwok \textit{et al.}, in which Ohmic dissipation was higher in current running in the direction perpendicular to the splay plane, suggesting that $\Jcm^{||\mathrm{ splay}}$ is larger than $\Jcm^{\perp\mathrm{ splay}}$ \cite{PhysRevLett.79.4258, PhysRevB.58.14594}.  Furthermore, the anisotropy in the \Jc was found to be pronounced at higher fields, suggestive of the fact that occupation of vortices in the columnar tracks heavily influence the pinning characteristics \cite{PhysRevLett.79.4258}. In compliment to transport measurements, MO images of \DBCO crystals with splayed columnar defects conjointly indicated that $\Jcm^{||\mathrm{ splay}} > \Jcm^{\perp\mathrm{ splay}}$ at high fields with anisotropy increasing at larger fields \cite{PhysRevB.51.16358}, and inversion in anisotropy where $\Jcm^{||\mathrm{ splay}} < \Jcm^{\perp\mathrm{ splay}}$ at low fields \cite{PhysRevB.51.16358}.  
%Moreover, the inversion in anisotropy was claimed to have been observed via MO imaging in \DBCO single crystals while absent in \YBCO single crystals.  The claim in the inversion in the anisotropy revealed through MO is unsettled due to the lack of clarity in the d-lines.

Yet, no observation has been ever made on IBS systems.  Hence, to confirm the existence of an anisotropic \Jc in \BaKx$\hspace{-1 pt}$, the spatial distribution of penetrated flux was observed through MO imaging at the remnant state.  Fig. \ref{M05}(b)-(d) illustrates the remnant state MO images after sweeping from 1 kOe back to zero field, at a temperature of 1 K below \Tc ($\approx 37$ K) in \BaKx with splayed columnar defects of $\pm 5 \Deg$, $\pm 10 \Deg$, and $\pm 15 \Deg$, respectively.  Just below $\Tcm$, at 37 K, the flux enters the center of the sample, forming a critical state, thereby leaving a double Y-shaped current discontinuity line. 

Unlike the isotropic case, as shown on the pristine crystal (Fig.\ref{M05}(a)) with a discontinuity line of $\approx 45\Deg$ angle with respect to the sample edge, there is anisotropy in the \Jc as evident from the skewed ``double Y'' discontinuity line, which appears as a consequence of the continuity condition \cite{:/content/aip/journal/apl/55/3/10.1063/1.102387, PhysRevB.53.2257}.  It is noteworthy that even for a small splay angle of $\pm 5 \Deg$, a remarkable anisotropy is observed.  Consistent to \YBCO and \DBCO with bimodal splayed columnar defects, we can confirm  $\Jcm^{||\mathrm{ splay}} > \Jcm^{\perp\mathrm{ splay}}$ for Ba$_{1-x}$K$_x$Fe$_2$As$_2$.  Similar trends in the anisotropy were also observed through MO imaging in crystals with $\pm 10 \Deg$ and $\pm 15 \Deg$ splay angles (Fig. \ref{M05}(c) and (d)).  For further quantitative analysis of the \Jc anisotropy, line profiles of the flux density along white (red) dashed lines in Figs. \ref{M05}(a), (b), (c), and (d) are shown in Figs. \ref{M15} (a)((b)), (c)((d)), (e)((f)), and (g)((h)), respectively.  Clearly, the distances of the flux peaks in the discontinuity lines from the sample edge are not equal.  Comparing the ratio of the distance of the discontinuity lines from the sample edge along and perpendicular to the splay, it is clear that the anisotropy increases with increasing splay angle.  While the anisotropy of the $\Jcm$ ($\zeta = \Jcm^{||\mathrm{ splay}} / \Jcm^{\perp\mathrm{ splay}}$) for the pristine crystal is $\zeta = 1$, $\zeta = 1.79$ for  $\pm 5 \Deg$, $\zeta = 2.63$ for $\pm 10 \Deg$, and $\zeta = 4.17$ for $\pm 15 \Deg$ splayed columnar defects (Fig. \ref{aniso}).  Hence, the \Jc in the direction of the splay plane has a value much larger than the \Jc in the perpendicular direction.  

Schuster \textit{et al.} suggest that the anisotropy in the \Jc is due to differences in activation barrier as a result of distinct kink nucleation process across and in the same direction of the splay: for $F_L^{\perp\mathrm{ splay}}$ ($\Jcm^{||\mathrm{ splay}}$), vortex motion is controlled by zig-zag type kinks, whereas for  $F_L^{||\mathrm{ splay}}$ ($\Jcm^{\perp\mathrm{ splay}}$), vortex motion is manifested by double-kinks \cite{PhysRevLett.79.4258}. L\'opez \textit{et al.} further this argument by advocating that the vortex structure is associated with the anisotropic dissipation in the superconductor.  The forced entanglement of vortices due to splayed columnar defects is effective only when vortices maintain $c$-axis coherence.  When the $c$-axis coherence of vortices is lost, vortices lose its identity as a line and are torn apart into decoupled segments of vortices.  Such vortex coherence is lost due to thermal decoupling of vortices and in the advent of flux cutting \cite{PhysRevLett.67.3176, PhysRevB.50.10294}. In light of this argument, the anisotropy would be only present at small splay angles since flux cutting is difficult, forcing vortices to entangle.  At large splay angles, flux cutting could be achieved easily and flux entanglement would not occur.  From MO images obtained in this experiment, even in the low-field regime close to the self-field, the significant \Jc anisotropy indicates great degree of flux entanglement.  Even at large splay angle of $\pm 15 \Deg$, flux entanglement is observed. In stark contrast to such high degree of anisotropy detected in IBSs at low fields, \Jc anisotropy is almost nullified in the remnant state magnetization  among \YBCO and \DBCO single crystals \cite{PhysRevLett.79.4258,PhysRevB.51.16358}.  This implicitly suggests that vortex coherence in IBSs are more robust than that of cuprates, as there is smaller anisotropy in coherence length amongst IBSs. 
%%%%%%%%%%%%%%%%%%%%%%%%%%%%%%%%%%%%%%%%%%%%%%%%
\subsection{Global critical current density}

  The relationship between the anisotropic critical current density and the magnetization for when $J_\mathrm{c}^{\parallel\mathrm{splay}}/J_\mathrm{c}^{\perp\mathrm{splay}}>a/b$, where $a$ and $b$ are the dimensions of the crystal, is given by 
\begin{align}
\Delta M = \frac{J_\mathrm{c}^{\perp\mathrm{splay}}a}{20}\bigg(1-\frac{a}{3b}\frac{J_\mathrm{c}^{\perp\mathrm{splay}}}{J_\mathrm{c}^{\parallel\mathrm{splay}}}\bigg)
\end{align}
\cite{:/content/aip/journal/apl/55/3/10.1063/1.102387}.  Although the two components of $\Jcm$ were decomposed through MO imaging at temperatures close to $\Tcm$ at the remnant state, determining the value of the individual $\Jcm$ components at higher fields cannot be performed by this method due to saturation of the Faraday rotation of the garnet indicator film.  Another method is through transport measurements.  Yet, this method would require applying a large current on the sample, or alternatively preparing a thin sample, which are both technically difficult.  Hence, we build our discussion based on global magnetization measurements, and calculate the overall $\Jcm$ given by the isotropic Bean's model, and compare the values between different splay angles, as done in Ref. \cite{PhysRevLett.76.2563}.
%In order to calculate the $\Jcm$, the Bean model for a sample with rectangular cross-section was employed.
  
In the conventional method, the width of the hysteresis loop $\Delta M$ which is the difference between $M $ sweeping down field and then back up field is used. However, since the self-field is significant, the return branch will cause a non-negligible effect on the calculation of $\Jcm$. Hence, instead,  the reversible linear background was first obtained through calculating the average of the magnetization of the second and the third quadrant.  This linear background component was subtracted from the raw data so that the hysteresis is virtually an even function, $M(H)=M(-H)$.  This allows for the calculation of the  $\Jcm$ from the magnetization of the second quadrant of the magnetic hysteresis using the extended isotropic Bean model,
\begin{align}
\Jcm = \frac{40 M}{a(1-a/3b)}
%\Jcm = \frac{40 M}{a(1-a/3b)} =  \frac{20 \Delta M}{a(1-a/3b)} 
%M = \frac{\Jcm}{40}\bigg(1-\frac{a}{3b}\bigg)
\end{align}
.  The error of $\Jcm$ due to the deviation of $M(H)$ from an ideal even function is estimated to be less than 8$\%.$

Fig. \ref{JcH} displays the $\Jcm(H)$ calculated from magnetic hysteresis loops.  As indicated in Fig. \ref{JcH}(a), the self-field $\Jcm$ of pristine crystals at 2 K exhibits a value of 2.6 MA/cm$^2$, consistent with other reports \cite{Ohtake201547, 0953-2048-28-8-085003}, while irradiated samples reveal a $\Jcm$ over 10 MA/cm$^2$, signifying substantial increase in flux pinning with incorporation of columnar defects. We see that for the case of parallel defects, there is a significant increase in the $\Jcm$, exhibiting a typical monotonic decrease with increasing field.  
% It is noteworthy that no dip-structures at fields below the self-field is seen as in the case for optimally doped Ba(Fe$_{1-x}$Co$_x$)$_2$As$_2$ irradiated with the same dose of $^{238}$U ions \cite{Yagyuda2011790}. Theoretically, in a thin superconductor, when the anisotropy in the coherence length between the $c$-direction and the $ab$-direction is pronounced, the dip structure at self-field is asserted to bud into a peak \cite{PhysRevB.62.6800}.

Figs. \ref{JcH}(c)-(f) indicate the $\Jcm$ as a function of magnetic field at various splay angles ranging from $\pm 5 \Deg$ to  $\pm 20 \Deg$. To compare the effects of the splayed columnar defects, we compare the value of $\Jcm $ at 2 K under self-field.  
%, where anisotropy is less significant.  
Given that the samples are approximate squares (i.e. $a \approx b$), the two components differ from the overall $\Jcm$ by a factor
\begin{align}
J_\mathrm{c}^{\perp\mathrm{splay}} &= \frac{2}{3-1/\zeta}J_\mathrm{c}\\
J_\mathrm{c}^{\parallel\mathrm{splay}} &= \frac{2\zeta}{3-1/\zeta}J_\mathrm{c}.
\end{align}
%However, as reported in  \YBCO and \DBCO single crystals, at low temperatures and low field regimes, the anisotropy in a system with bimodal splayed columnar defects diminishes to unity, indicating an isotropic $\Jcm$ such that $\zeta\approx1$ \cite{PhysRevLett.79.4258,PhysRevB.51.16358}.  Therefore, the value of  $\Jcm$ at 2 K under self-field would be equal to the individual $\Jcm$ components (i.e. $\Jcm$(2 K, 0 T) $\approx J_\mathrm{c}^{\perp\mathrm{splay}}$(2 K, 0 T) $\approx J_\mathrm{c}^{\parallel\mathrm{splay}}$(2 K, 0 T) ).  
As shown in \YBCO and \DBCO single crystals, the anisotropy $\zeta$ in a system with bimodal splayed columnar defects has an intricate dependence on the field and temperature \cite{PhysRevLett.79.4258,PhysRevB.51.16358}.  In this investigation, to circumvent such complexities, we strictly limit our discussion on the average critical current density, $\Jcm$, obtained through magnetization measurements.

At 2K under self-field, for the case of those irradiated with parallel columnar defects (Fig. \ref{JcH}(b)), the $\Jcm$ exhibits a value of 13.9 MA/cm$^2$.  The value of \Jc obtained in this investigation for parallel columnar defects is comparable to that in the previous report of \Jc in 2.6 GeV U irradiated optimal \BaKx \cite{Ohtake201547}. Strikingly, the $\Jcm$ of samples irradiated with a splay angle of $\pm 5 \Deg$ at 2 K displays a value of 19.5 MA/cm$^2$, exceeding those with parallel columnar defects and larger splay angles under all field ranges (Fig. \ref{angle}). Moreover, it is clear that splay angles larger than $\pm 5 \Deg$ exhibit a lower $\Jcm$, suggesting that the effects of vortex-field misalignment outperforms the enhancement effect of splayed defects when the tilt angle of columnar defects increases.  This result is consistent with \YBCO single crystals in Ref. \cite{PhysRevLett.76.2563}, where $\pm 5 \Deg$ was reported to be the optimal splay angle with decreasing $\Jcm$ at higher splay angles.  

Not to mention, amongst those with splayed columnar defects, there is an apparent non-monotonicity in the $\Jcm$ with increasing field  (Fig. \ref{JcH}).  Intuitively, the \Jc should monotonically decrease with increasing fields due to larger driving force to pull the vortex from its pinning center.  For the case of the pristine sample, the highest $\Jcm$ at all temperature regimes reside at remnant magnetization.  Upon inducing parallel columnar defects, a peak-like behavior appears at intermediate temperatures and at low fields. Such behavior can be inferred to originate from the curvature of vortices around the self-field which induces depinning \cite{0953-2048-25-8-084008}. However, in those with splayed columnar defects, a much larger and broader peak occurs at higher fields.  Since the self-field effect do not occur at large fields, the non-monotonic behavior arising in a system with splayed columnar defects differs from that with parallel defects.
\begin{figure*}[t]
  \begin{center}
\includegraphics[width=0.8\textwidth]{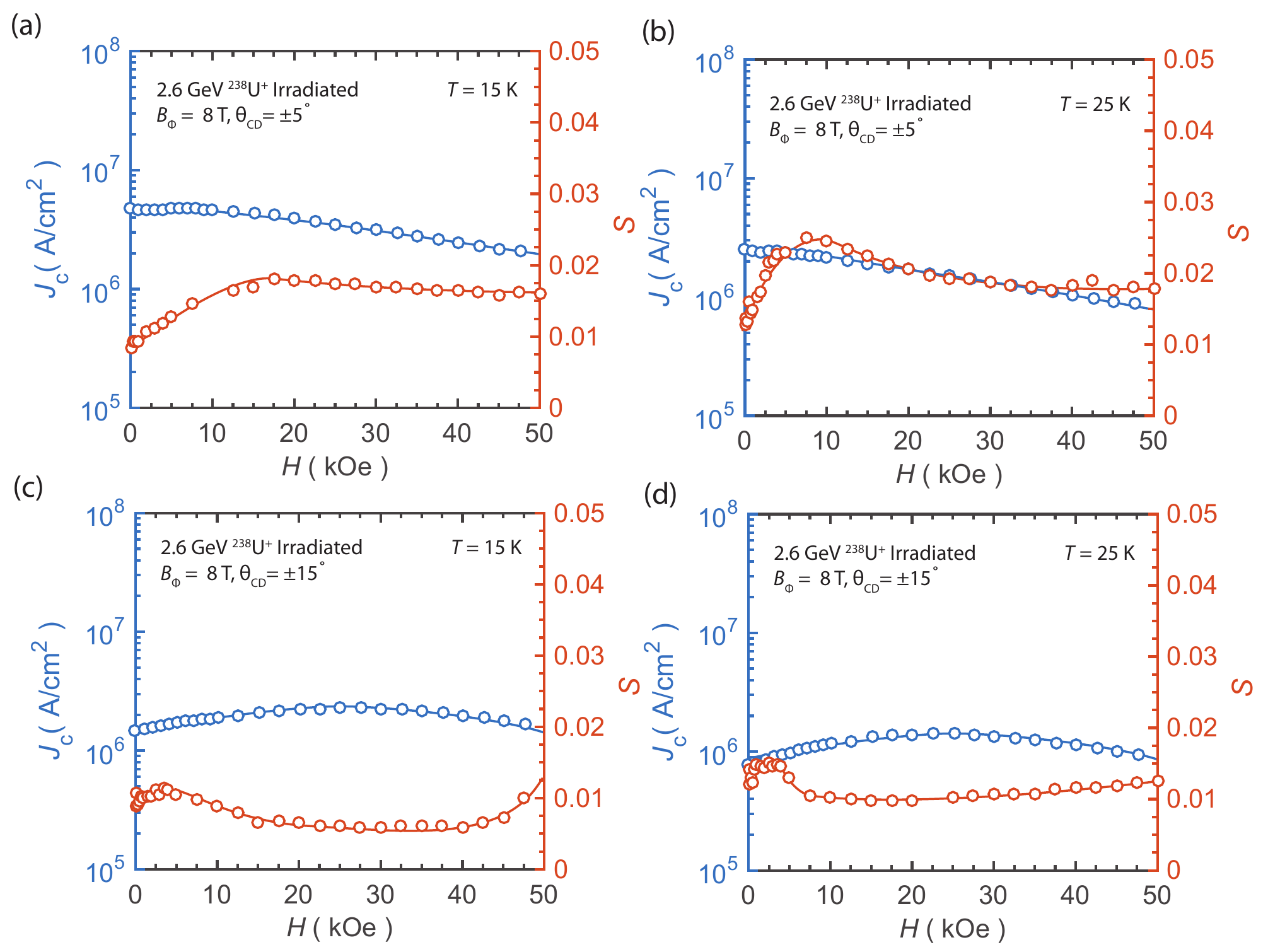}% Here is how to import EPS art
  \end{center}
%\vspace{30 pt}
\caption{ {The field dependence of the normalized relaxation rate of \BaKx with splay of $\pm 5 \Deg$ at (a) 15 K and (b) 25 K.  Similarly, $S$ for  splay of $\pm 15 \Deg$ at (c) 15 K and (d) 25 K.} }
\label{relax1}
\end{figure*}

\begin{figure}[t]
  \begin{center}
\includegraphics[width=8.5 cm]{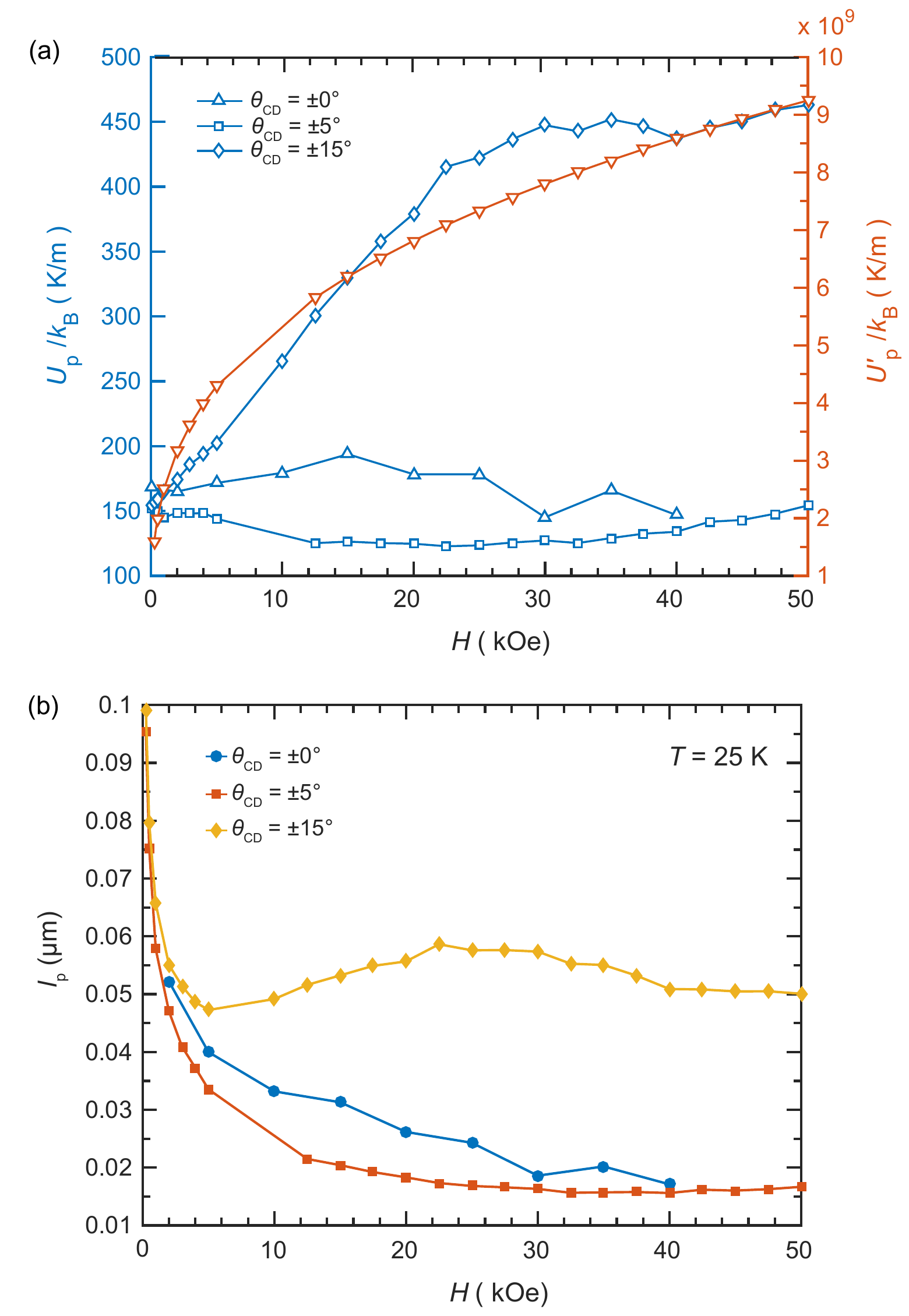}% Here is how to import EPS art
  \end{center}
\caption{ {(a) The field dependence of $U_\mathrm{p}$ and  $U_\mathrm{p}'$ at 25 K for samples with splayed columnar defects with angles $\pm 0 \Deg$, $\pm 5 \Deg$, and $\pm 15 \Deg$. (b) The field dependence of $l_\mathrm{p}$ calculated from $U_\mathrm{p}$ and  $U_\mathrm{p}'$.} }
\label{lp}
\end{figure}

\begin{figure}[t]
  \begin{center}
\includegraphics[width=8 cm]{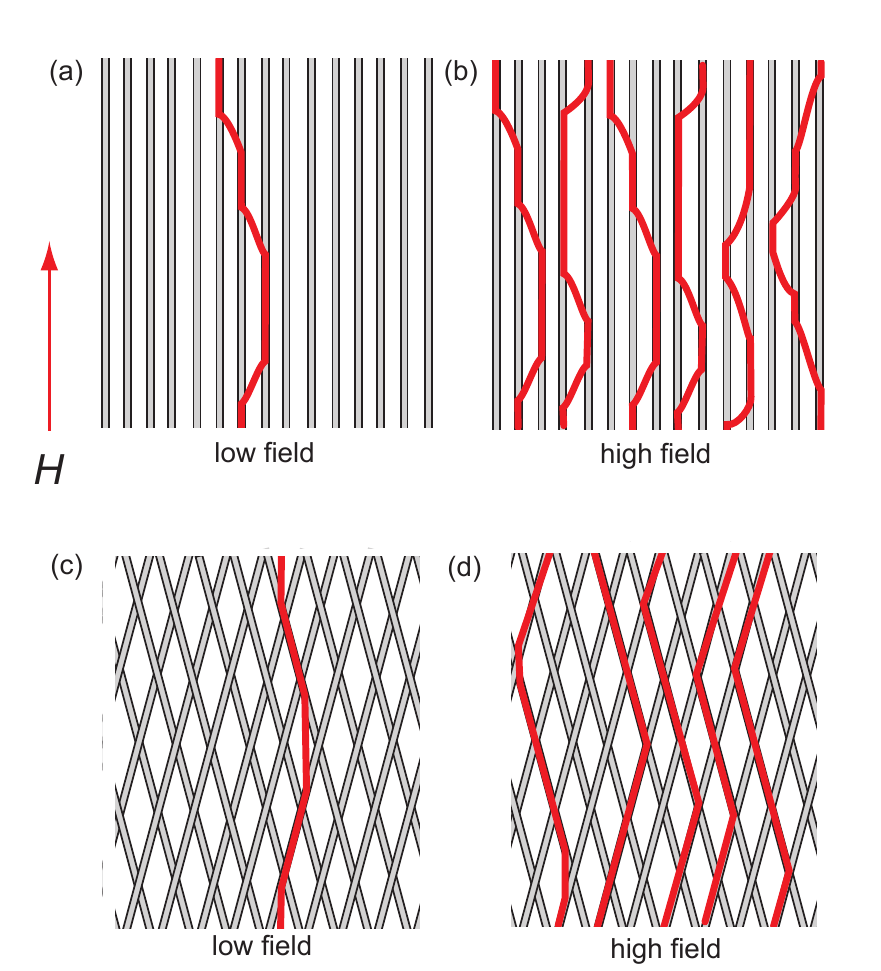}% Here is how to import EPS art
  \end{center}
\caption{ {Schematics of the vortex structure in a Bose glass phase at (a) low fields and at (b) high fields at a temperature of 25 K.  The vortex structure in a splayed glass phase (c) at low fields and (d) at high fields. } }
\label{vortex2}
\end{figure}

\begin{figure}[h!]
  \begin{center}
\includegraphics[width=8 cm]{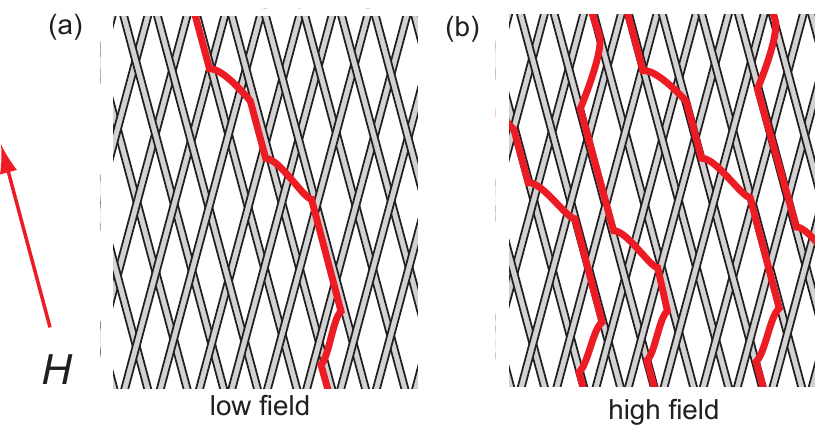}% Here is how to import EPS art
  \end{center}
\caption{ {Schematics of the vortex structure in a splayed glass phase (a) at low fields and (b) at high fields when the field is applied in the direction that corresponds to one of the modes of the columnar defects. } }
\label{vortex3}
\end{figure}

\subsection{Effects of Tilted Fields}
To further investigate the effects of vortex entanglement in the magnetization amongst systems with splayed columnar defects, the magnetization was measured while tilting the angle of the field in the direction of the splay plane (Fig. \ref{tilt1}).  Since the magnetization is detected only in the direction of the field, the actual magnetization of the sample is compensated by multiplying by a factor of $1/\cos(\theta_\mathrm{H})$. At the lowest temperature, the magnetization is independent of the angle of the tilted field as indicated by the flat field-angle ($\theta_\mathrm{H}$) dependence exhibited in Fig. \ref{tilt1}(d).  The effect of tilted field becomes more prominent at higher temperatures, where vortices are less rigid.  By tilting the field closer to one of the two modes of columnar defects, surprisingly, the non-monotonic behavior is completely eradicated, with a hysteresis reminiscent of that observed in crystals with parallel columnar defects as shown in Fig. \ref{JcH}(b).  Remarkably, the \Jc is significantly higher in the case when $\theta_H || c$ than $\theta_\mathrm{H} = \theta_{\mathrm{CD}}$ at high fields, similar to the behavior of the $\Jcm$ angluar dependence of 270 MeV Xe irradiated $RE$Ba$_2$Cu$_3$O$_y$ coated conductors \cite{Sueyoshi201453, 0953-2048-29-10-105006}.  

We compare this to the case with a tilted (single-mode) columnar defect system of $\theta_\mathrm{CD} = 5 \Deg$.  As shown in Fig. \ref{tilt2}, the non-monotonic behavior seen in splayed systems is absent.  Hence, clearly the non-monotonicity in the field dependence of \Jc is characteristic to systems with splayed columnar defects.  Although the \Jc is almost independent of $\theta_\mathrm{H}$ at low temperatures, the differences becomes more prominent at higher temperatures.  At high fields ($2$ T or larger), where self-field effect does not play a role, the highest \Jc occurs when $\theta_\mathrm{H} = \theta_{\mathrm{CD}}$.  This is due to the energetically stable flux composition in which it is aligned to the external magnetic field.

\subsection{Magnetic Relaxation Rate}
Amongst \YBCO single crystal, one intriguing feature is while splay enhances the critical current density especially in those with small splay angles, flux creep is reported to be promoted upon incorporation of splayed columnar defects \cite{PhysRevLett.76.2563, PhysRevLett.78.3181}.  Fig. \ref{relax1} illustrate the field dependence of the normalized magnetic relaxation rate $S$ defined by

\begin{align}
S = \bigg|\frac{d\ln(M)}{d\ln(t)}\bigg|,
\end{align}

\noindent for splay angles of $\pm 5 \Deg$ and $\pm 15 \Deg$ at 25 K.  Clearly, the field dependence of $S$ in splayed systems are distinct from those with parallel columnar defects (2.6 GeV U ions with a dose of 8 T) \cite{Ohtake201547}.  
%For the case of parallel defects at 5 K, in low fields, there is a peak in the relaxation rate, corresponding to the depinning due to the self-field effect with a value of $S \approx 0.03$.  At higher fields, the relaxation rate flattens to a value of $S \approx 0.01$.  Similar trend can be observed in the case of splayed columnar defects with $\pm 15 \Deg$ as shown in Fig. \ref{relax1}(b).  However, with $\pm 5 \Deg$, at low fields, there are two peaks present in the field dependence of the relaxation rate (Fig. \ref{relax1}(b)). The peak on the lower field is conventionally interpreted to arise owing to the self-field effect.  Despite the presence of an anomalous secondary peak residing at a field of $1.25$ T, no particular change in the critical current density occurs at this field range.  
%Moreover, 
It is noteworthy that at higher fields, relaxation rate in sample with $\theta_{\mathrm{CD}}=\pm 5 \Deg$ is higher than that with $\theta_\mathrm{CD}=\pm 15 \Deg$.  Most importantly, the field dependence of the critical current densities are depicted to be mirror-images of the field dependence of the relaxation rate.  The local maxima in the \Jc for  $\pm 15 \Deg$ corresponds to the local minima in $S$.  Such mirror-image correspondence between $S$ and \Jc has been observed in both cuprates and IBSs \cite{PhysRevLett.69.2280, PhysRevB.78.214515}.  
%However, for this case, we attribute the local minima in $S$ to the maxima in the $\Jcm$.  

\section{Discussions}

Up to this point, we have observed an anomalous behavior in vortex pinning and vortex dynamics in those with splayed columnar defects  at intermediate temperatures.  Moreover, remarkably, when tilting the field along the splay plane, the anomalous peak in the magnetization is eliminated.  In order to reveal such strange behavior, we expand our discussion on the vortex structure amongst systems with splayed columnar defects in IBSs.

We begin by considering the angular behavior of the pinning energy per unit length ($U_\mathrm{p}'$) in a system with columnar defects %(Eq. \eqref{up})
 \cite{PhysRevB.48.10487, PhysRevB.56.913},
\begin{align}
U_\mathrm{p}' \approx \varepsilon_0\bigg(\frac{2k_\mathrm{B}T\tan (\theta_\mathrm{acc})}{\varepsilon_0a_0}\bigg)^{2/3},
\label{up}
\end{align}
where $\varepsilon_0=(\Phi_0/4\pi\lambda_{ab})^2$ is the vortex line energy, $a_0=\sqrt{\Phi_0/B}$ is the average inter-vortex spacing, and $\theta_\mathrm{acc}$ is the accommodation angle. 
%Using $\lambda(0) = 200$ nm \cite{PhysRevLett.101.107004}, along with the temperature dependence $\lambda(T)  = \lambda(0)(1-T/T_\mathrm{c})^{-1/2}$, we obtain $\lambda(25 \hspace{1 pt}\mathrm{K}) = 340$ nm, which is used to calculate $\varepsilon_0$. 
The vortex accommodation angle is obtained from the vortex lock-in angle using 
%Eq. \eqref{tL}
the following relationship
\begin{align}
\theta_\mathrm{L} = \frac{4\pi\varepsilon_l}{\Phi_0B}\theta_\mathrm{acc}
\label{tL}
\end{align}
 \cite{PhysRevB.48.13060, RevModPhys.66.1125, bennemann2012physics} in which $\varepsilon_l =\varepsilon_0\ln(\kappa)$ is the line tension, with $\kappa = \lambda/\xi$ being the Ginzburg-Landau parameter. For this case, we use the experimentally obtained lock-in angle ($\theta_\mathrm{L}$) reported in Ref. \cite{PhysRevB.89.024508}.   We note that although the lock-in angle obtained in Ref.  \cite{PhysRevB.89.024508} is that of  2.6 GeV U irradiated Ba(Co$_{1-x}$Fe$_x$)$_2$As$_2$, since \BaKx has similar anisotropy, we estimate that the lock-in angle should be no different for both cases.  Thus, using $\xi(0) = 1.2$ nm \cite{PhysRevB.79.094505}, and $\lambda(0) = 200$ nm \cite{PhysRevLett.101.107004}, along with the temperature dependences $\xi(T)  = \xi(0)(1-T/T_\mathrm{c})^{-1/2}$, $\lambda(T)  = \lambda(0)(1-T/T_\mathrm{c})^{-1/2}$, we obtain  $\xi(25 \hspace{1 pt}\mathrm{K}) = 2.1$ nm and $\lambda(25 \hspace{1 pt}\mathrm{K}) = 340$ nm, allowing us to acquire the a value of $\theta_\mathrm{acc} = 41.4^{\circ}$, which is field-independent.
%  In Eq. \eqref{tL} the line tension is $\varepsilon_l =\varepsilon_0\ln(\kappa)$, where $\kappa = \lambda/\xi$ is the Ginzburg-Landau parameter. Thus using $\xi(0) = 1.2$ nm \cite{PhysRevB.79.094505}, and $\lambda(0) = 200$ nm \cite{PhysRevLett.101.107004}, along with the temperature dependences $\xi(T)  = \xi(0)(1-T/T_\mathrm{c})^{-1/2}$, $\lambda(T)  = \lambda(0)(1-T/T_\mathrm{c})^{-1/2}$, we obtain  $\xi(25 \hspace{1 pt}\mathrm{K}) = 2.1$ nm and $\lambda(25 \hspace{1 pt}\mathrm{K}) = 340$ nm, which are used to calculate $U_\mathrm{p}'$ (Fig. \ref{lp}(a)). 

We compare this value to the actual pinning energy in the system with splayed columnar defects $U_\mathrm{p}$ by considering the  inverse power-law barrier proposed by Feigelfman
\begin{align}
U = U_\mathrm{p}\bigg(\bigg(\frac{J_\mathrm{c0}}{J}\bigg)^\mu-1\bigg)
\label{Uj}
\end{align}
\cite{PhysRevLett.63.2303}.  Here, $U$ is the effective activation energy, $J$ is the current density, $J_\mathrm{c0}$ is the critical current density required to nullify the activation energy, and $\mu$ is the glassy exponent
% which gives information on the vortex bundle size within the collective creep theory
%.  The parameter of interest, $U_p$, is obtained through considering a temperature dependence $U_\mathrm{p} = U_\mathrm{p0}(1-(T/T_\mathrm{c})^2)^{3/2}$, and using  Eq. \eqref{Uj} to fit experimentally obtained magnetic relaxation data scaled by the Maley's relationship
.  The value of $U_p$, is obtained using  Eq. \eqref{Uj} to fit the experimentally obtained magnetic relaxation data scaled by Maley's relationship
%Anderson-Kim activation energy \cite{PhysRevLett.9.309, RevModPhys.68.911}.
\begin{align}
U = -k_\mathrm{B}T\bigg(\ln\bigg(\frac{dM}{dt}\bigg) - C\bigg)
\label{upp}
\end{align}
through a non-linear least squares method, where $C$ in Eq.\eqref{upp} is an arbitrary constant, which we fix with the value $C=30$ for samples with splay angles of $\pm 5 \Deg$ and $\pm 15 \Deg$, and $C=20$ for sample with parallel columnar defects \cite{PhysRevB.42.2639}.  Upon fitting, we consider the temperature dependence $U_\mathrm{p} = U_\mathrm{p0}(1-(T/T_\mathrm{c})^2)^{3/2}$ \cite{PhysRevB.86.094527}, and fix the glassy exponent $\mu = 7/9$ at the large vortex bundle regime (i.e. Larkin lengths are larger than the penetration depth) since the vortices are expected to be highly correlated with such high degree of disorder.  Thus we obtain the field-dependence of the activation barrier as exhibited in Fig. \ref{lp}(a).

From $U_\mathrm{p}$ and $U_\mathrm{p}'$, we can obtain the effective length of the vortex segment trapped in the columnar defect ($l_\mathrm{p}$)
\begin{align}
l_\mathrm{p} \approx U_\mathrm{p}/U_\mathrm{p}'
\end{align}

\cite{PhysRevB.58.14594}.  Fig. \ref{lp}(b) illustrates the evolution of $l_\mathrm{p}$ with increasing field at a temperature of 25 K.  It becomes evident that for the case when columnar defects are parallel ($\theta_\mathrm{CD} = \pm0 \Deg$),  $l_\mathrm{p}$ decreases with increasing field and becomes field-independent at a value of $l_\mathrm{p} \approx 0.02$ $\mu$m.  Similarly, for the case of $\theta_\mathrm{CD} = \pm5 \Deg$, there is a steady decrease in $l_\mathrm{p}$ upon an increase in the field, reaching a value of $\approx 0.02$ $\mu$m at a field of 50 kOe.  Even more strikingly, for $\theta_\mathrm{CD}  =\pm15 \Deg$, a less rapid decrease in $l_\mathrm{p}$ is evident, where $l_\mathrm{p} \approx 0.05$ $\mu$m at a field of 50 kOe.  Hence, amongst a splayed glass phase, there is an apparent robustness in the effective length of the vortex segment pinned to the columnar defect with an increase in field.  

From such observation, we infer that the vortex structure with parallel columnar defects at low fields are essentially linear with certain number of kinks that reach out to neighboring columns due to thermal fluctuations as illustrated in Fig. \ref{vortex2}(a).  As the field increases, the overall vortex density increases with significant inter-vortex interaction. Yet, the passive change in $l_\mathrm{p}$ value entails that the vortex structure remains largely unaltered (Fig. \ref{vortex2}(b)). 

 For the case with splayed columnar defects at low fields, the vortices are fundamentally trapped in the columnar defects with some thermally activated kinks as depicted in Fig. \ref{vortex2}(c).  Upon increasing the field, the vortices are accommodated into the defects, forming a ``zig-zag" configuration, as reflected in the increase in  $l_\mathrm{p}$.  Since a higher degree of pinned vortex length results in a stronger pinning, we suggest that such change in the vortex structure could be highly related to the non-monotonic field dependence of $\Jcm$.
%  This raises the question: what drives such change in the vortex structure ?
 
In regards to this framework, we must explicate why the non-monotonic field dependence is eliminated when the field is applied in the direction that corresponds to one of the modes of the bimodal splay.  As shown in Fig. \ref{vortex3}(a), since one of the modes of the splay is already in-line with the field, it is anticipated that the vortex should be linear, as it is the most energetically stable configuration. Therefore, even in high-field regimes, the vortex configuration remains unchanged (Fig. \ref{vortex2}(b)), as with the case with parallel columnar defects, thereby resulting in a conventional monotonic field dependence of $\Jcm$.
 
 A similar phenomena of \Jc  enhancement has been reported to occur in heavy-ion irradiated 
\YBCO single crystals \cite{PhysRevLett.67.648, PhysRevLett.86.5144}.  Such phenomena arise at a field of $1/5\sim1/3B_\Phi$ which corresponds to the field range of the peak seen in this case.   Although the reported \YBCO crystals were irradiated parallel to the $c$-axis by heavy ions, cross-sectional TEM images reveal naturally induced splayed columnar defects \cite{PhysRevLett.67.648}.  Moreover, consistent to U-irradiated IBSs in this investigation, only when the field is in the same direction of the $c$-axis, the non-monotonic behavior ensues, while disappearing when tilted in an angle.  To explicate this behavior, it has been suggested that the non-monotonicity of $\Jcm$ emanates from increased inter-vortex repulsion which increases the vortex trapping rate by columnar defects.  As a result of amplified vortex trapping, increased interlayer coupling coherence of vortices is achieved. Indeed, enhancement of interlayer coherence has been confirmed experimentally through Josephson plasma resonance measurements in heavy-ion irradiated Bi$_2$Sr$_2$CaCu$_2$O$_{8+y}$, signifying the enhancement of vortex trapping.  The similarities between the two systems with splayed columnar defects suggests that  the non-monotonic field dependence of \Jc  in this framework is possibly a universal behavior that does not only apply to IBSs.
\section{Conclusions}

Through this investigation, we have initially revealed four main observations. (1) By introducing a bimodal splay through U-irradiation,  \BaKx single crystals exhibit a highly anisotropic \Jc with even greater anisotropy with larger splay angles.  (2)  System with splay angle of $\pm 5 \Deg$, reveals an optimal \Jc with a high value of 19.5 MA$/$cm$^2$.  (3) Thirdly, and most importantly, amongst a splayed glass phase, an anomalous non-monotonic field dependence of \Jc and $S$ arises. (4) Last but not the least, through tilting the field so that the field is aligned to one of the two modes of splay, the non-monotonic $\Jcm$ dependence is strangely eradicated. 

In order to interpret such salient non-monotonicity in the field-dependence of $\Jcm$, we examine the evolution of the effective length of vortex segment trapped in the columnar defect $l_\mathrm{p}$ with increasing magnetic field and reveal that systems with splayed columnar defects exhibit a larger value in $l_\mathrm{p}$ than that in those with parallel columnar defects.  The accommodation of vortices into columnar defects in splayed systems are reminiscent of the field-driven interlayer re-coupling transition behavior seen in heavy-ion irradiated cuprates.  Such reported phenomena and the one seen in this investigation are highly consistent, as they appear in similar field ranges.  However, there is an essential difference between the two such that while in cuprates, the $\Jcm$ non-monotonicity is seen in those with parallel columnar defects, we see that  the behavior is absent amongst IBSs with parallel columnar defects and only present in those with splayed columnar defects.  We discuss that the inherent disparity is due to differences in the strength of the vortex interlayer coupling and the defect morphology apparent in the two systems.

Finally, we reiterate the fact that the investigation presented here is based on analysis of the average in-plane $\Jcm$ rather than treating the individual $\Jcm$ components.  Further analysis on the effects on the anisotropy of $\Jcm$ induced by splayed columnar defects would further shed light into the complex vortex behavior in such systems. 
%Such zig-zag formation is also suggested by a peak in the normalized relaxation rate at low fields, indicating a high degree of kinks with pin-free segments.  However, with an increase in field, the relaxation rate decreases, indicating a decrease in the number of kinks, in other words, implying an accomodation of vortices into vortices, driven by enhanced vortex-vortex interactions. Moreover, through suppressing the entanglement of vortices so that the field and the columnar defects are aligned, there is absence of such anamolous peak in \Jc.
%Up to this point, we have revealed experimentally, four main points. Firstly, samples with splayed columnar defects exhibit a non-monotonic field dependence of $\Jcm$.  Secondly, through MO imaging we have confirmed the existence of $\Jcm$ anisotropy at low-fields, suggesting an entangled vortex phase.  Furthermore, through tilting the field so that the field is aligned to one of the two modes of splay, the non-monotonic $\Jcm$ dependence is eradicated.  Last but not the least, the peak in the $\Jcm$ is reflected in the local minima of the $S$.  Based on these points, we would like expand our discussion on the vortex structure amongst splayed systems in IBSs in order to explicate the observed anamolous non-monotonic field dependence of $\Jcm$.

\vspace{ 10 pt}
\section*{Acknowledgments}

This experiment was performed at RI Beam Factory operated by RIKEN Nishina Center and CNS, The University of Tokyo.  This work is partly supported by KAKENHI (17H01141) from JSPS.

%\bibliography{bibliography}% Produces the bibliography via BibTeX.

\end{document}